# Search for Stopped Gluinos in $pp$ collisions at $\sqrt{s}$ = 7 TeV

*Fedor Ratnikov for CMS Collaboration*[*] [1]

(1) *Karlsruhe Institute of Technology, 76128 Karlsruhe, Germany,* fedor.ratnikov@cern.ch
[*] *on leave from Institute for Theoretical and Experimental Physics, Moscow*

## Abstract

The results of the first search for long-lived gluinos produced in 7 TeV $pp$ collisions at the CERN Large Hadron Collider are presented. The search looks for evidence of long-lived particles that stop in the CMS detector and decay in the quiescent periods between beam crossings. In a dataset with a peak instantaneous luminosity of $1 \times 10^{32}$ cm$^{-2}$s$^{-1}$, an integrated luminosity of 10 pb$^{-1}$, and a search interval corresponding to 62 hours of LHC operation, no significant excess above background was observed. Limits at the 95% confidence level on gluino pair production over 13 orders of magnitude of gluino lifetime are set. For a mass difference $m_{\tilde{g}} - m_{\tilde{\chi}^0_1} > 100$ GeV$/c^2$, and assuming BR$(\tilde{g} \to g\tilde{\chi}^0_1) = 100\%$, $m_{\tilde{g}} < 370$ GeV$/c^2$ are excluded for lifetimes from 10 $\mu$s to 1000 s.

Many extensions of the standard model predict the existence of new heavy quasi-stable particles [1]. Such particles are present in some supersymmetric models [2, 3, 4], "hidden valley" scenarios [5], and grand-unified theories (GUTs), where the new particles decay through dimension five or six operators suppressed by the GUT scale [6]. Long-lived particles are also a hallmark of split supersymmetry [7], where the gluino ($\tilde{g}$) decay is suppressed due to the large gluino-squark mass splitting, from which the theory gets its name. Of these possibilities, the Compact Muon Solenoid (CMS) experiment is most sensitive to models like split supersymmetry where production proceeds via the strong interaction resulting in relatively large cross sections at the Large Hadron Collider (LHC) [8, 9, 10, 11]. For this reason, we have targeted the search at long-lived gluinos. Existing experimental constraints on the lifetime of such gluinos are weak [12, 13]; these gluinos may be stable on typical CMS experimental timescales. Lifetimes of $\mathcal{O}$(100–1000) seconds are especially interesting in cosmology since such decays would affect the primordial light element abundances, and could resolve the present discrepancy between the measured $^6$Li and $^7$Li abundances and those predicted by conventional big-bang nucleosynthesis [14, 15, 16].

If long-lived gluinos were produced at the LHC, they would hadronize into $\tilde{g}g, \tilde{g}q\bar{q}, \tilde{g}qqq$ states, collectively known as "R-hadrons" some of which would be charged, while others would be neutral [17, 18, 19]. Those that were charged would lose energy via ionization as they traverse the CMS detector. For slow R-hadrons, this energy loss would be sufficient to bring a significant fraction of the produced particles to rest inside the CMS detector volume [20]. These "stopped" R-hadrons may decay seconds, days, or even weeks later, resulting in a jet-like energy deposit in the CMS calorimeter. These decays will be out-of-time with respect to LHC collisions and may well occur at times when there are no collisions in CMS. The observation of such decays, in what should be a "quiet" detector except for an occasional cosmic ray, would be an unambiguous discovery of new physics.

A detailed description of the CMS experiment can be found elsewhere [21].

The 7 TeV center-of-mass $pp$ collision data analyzed in this paper were recorded by CMS in October 2010. This sample corresponds to 62 hours of trigger live-time during which data, corresponding to an integrated luminosity of 10 pb$^{-1}$, were recorded by CMS with a peak instantaneous luminosity of $1 \times 10^{32}$ cm$^{-2}$s$^{-1}$. In producing these data, the LHC was filled with up to 312 proton bunches per beam (out of a maximum of 2808).

We employed a dedicated trigger to search for de-cays of particles at times when there are no collisions. Special selection criteria are applied during data analysis to suppress contributions from bunch crossing preceeding or following the colliding bunch crossing, cosmics muons, beam halo muons, beam-gas interactions, and instrumental noise. Details of trigger and selections are described elsewhere [22, 23].

We have developed a custom, factorized simulation of gluino production, stopping, and decay to investigate the experimental signature of this atypical signal [29]. First, using PYTHIA [24] we generate gluino production at $\sqrt{s} = 7TeV$ TeV, and hadronize the produced gluinos into R-hadrons. A modified GEANT4 [25] then implements a "cloud model" of interactions R-hadrons with matter [26]. Using this simulation, the probability of a single R-hadron to stop in the CMS detector was determined to be $\approx 0.2$ for the explored gluino mass range. We also considered alternative, more pessimistic, models of R-hadronic interactions with matter. For electromagnetic interactions (EM) only, the CMS stopping probability is found to be $\approx 0.06$. Finally, with a "neutral R-baryon" model in which only R-mesons stop [27, 28] the stopping probability is $\approx 0.01$.

Next, we again use PYTHIA, to simulate the decay of an R-hadron at rest in the previously recorded stopping position, and GEANT to simulate the detector response to this decay. Finally, we use a specialized Monte Carlo simulation to determine how often the stopped gluino decay would occur during a triggerable beam gap.

The efficiency with which triggered events pass all selection criteria is estimated from the simulation to be 54% for a representative gluino decay signal ($m_{\tilde{g}} = 300$ GeV/$c^2$ and $m_{\tilde{\chi}_1^0} = 200$ GeV/$c^2$). The equivalent efficiency with respect to all stopped particles is 17% since a significant number of R-hadrons stop in uninstrumented regions of the CMS detector where their subsequent decay would not be observable. For any new physics model that predicts events with sufficient visible energy, $m_{\tilde{g}} - m_{\tilde{\chi}_1^0} > 100$ GeV/$c^2$, this efficiency does not change significantly.

The background rate is measured by combining measurements in the control sample with measurements of the rate in the search sample after omitting one selection criterion as described in [22, 23].

We do not observe a significant excess above expected background for any lifetime hypothesis in the search sample. The results of this counting experiment for different lifetime hypotheses are presented in Table 1. In the absence of any discernible signal, we proceed to set 95% confidence level (C.L.) limits over 13 orders of magnitude in gluino lifetime using a hybrid $CL_S$ method.

Table 1: *Results of counting experiments for selected values of* $\tau_{\tilde{g}}$. *Entries between* $1 \times 10^{-5}$ *and* $1 \times 10^{6}$ *s are identical and are suppressed from the table.*

| Lifetime [s] | Expected Background | Observed |
|---|---|---|
| $1 \times 10^{-7}$ | $0.8 \pm 0.2 \pm 0.2$ | 2 |
| $1 \times 10^{-6}$ | $1.9 \pm 0.4 \pm 0.5$ | 3 |
| $1 \times 10^{-5}$ | $4.9 \pm 1.0 \pm 1.3$ | 5 |
| $1 \times 10^{6}$ | $4.9 \pm 1.0 \pm 1.3$ | 5 |

In Fig. 1 we show the 95% C.L. limit on $\sigma(pp \to \tilde{g}\tilde{g}) \times BR(\tilde{g} \to g\tilde{\chi}_1^0)$ for a mass difference $m_{\tilde{g}} - m_{\tilde{\chi}_1^0} > 100$ GeV/$c^2$. The error bands include statistical and systematic uncertainties. With the horizontal line in Fig. 1 we show a recent NLO+NLL calculation of the cross section at $\sqrt{s} = 7$ TeV for $m_{\tilde{g}} = 300$ GeV/$c^2$ from the authors of Ref. [11]. To illustrate the effect of the stopping probability uncertainty, we also present two other limits for different models of interaction R-hadrons with material. Assuming the cloud model for the interaction of R-hadrons with matter, and assuming $\mathrm{BR}(\tilde{g} \to g\tilde{\chi}_1^0) = 100\%$, we are able to exclude lifetimes from 75 ns to $3 \times 10^5$ s for $m_{\tilde{g}} = 300$ GeV/$c^2$ with the counting experiment. Finally, we present the result as a function of the gluino mass in Fig. 2. Under the same assumptions as for the cross section limit, we exclude $m_{\tilde{g}} < 370$ GeV/$c^2$ for lifetimes between 10 $\mu$s and 1000 s. If we assume the EM only model for R-hadronic interactions with matter in order to compare with what was done in Ref. [12], this exclusion becomes $m_{\tilde{g}} < 302$ GeV/$c^2$.

We also perform a time-profile analysis. Whereas, for short lifetimes, a signal from a stopped gluino decay is correlated in time with the collisions, backgrounds are flat in time. Since the signal and background have very different time profiles, it is possible to extract both their contributions by analyzing the distribution of the observed events in time.

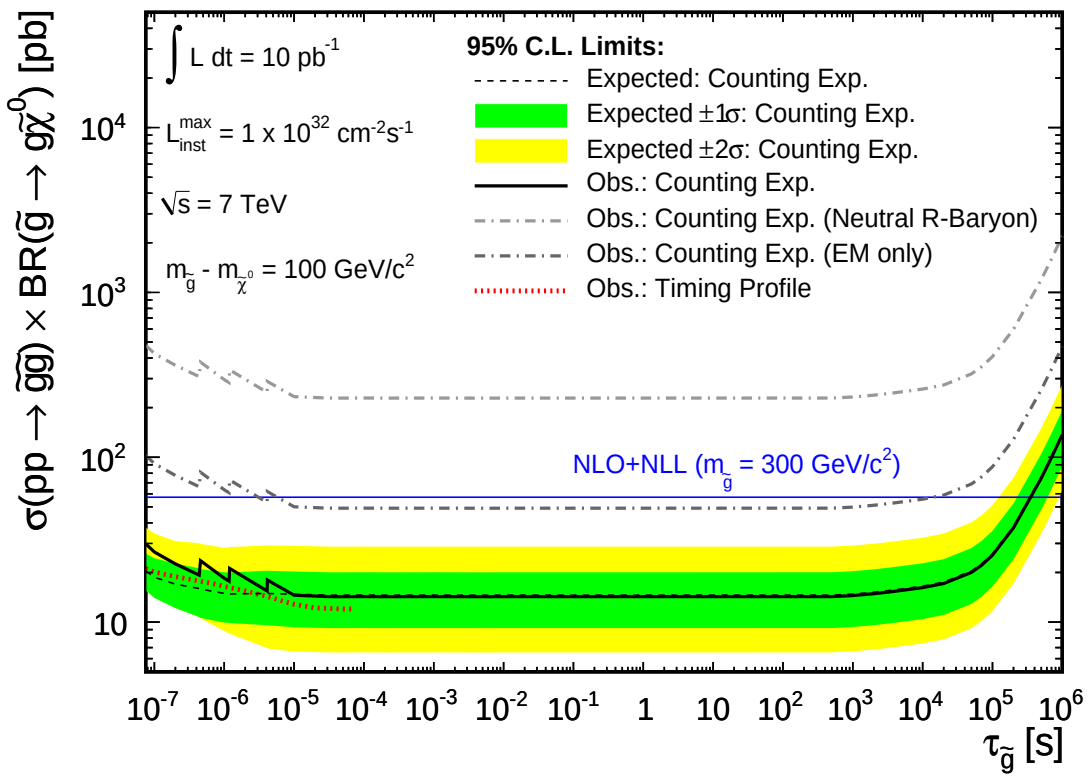

***Fig. 1:** Expected and observed 95% C.L. limits on gluino pair production cross section times branching fraction using the "cloud model" of R-hadron interactions as a function of gluino lifetime from both the counting experiment and the time-profile analysis. Observed 95% C.L. limits on the gluino cross section for alternative R-hadron interaction models are also presented.*

We build a probability density function (PDF) for the gluino decay signal as a function of time for a given gluino lifetime hypothesis and the actual times of LHC beam crossings as recorded in our data. Figure 3 shows an example of such a PDF for a gluino lifetime of 1 $\mu$s; the in-orbit positions of 2 observed events in the subset of our data that were recorded during an LHC fill with 140 colliding bunches are overlaid. We limit the range of lifetime hypotheses considered for this time-profile analysis to 75 ns to 100 $\mu$s such that the gluino lifetime is not much longer than the orbit period. For each lifetime hypothesis we build a corresponding signal time profile, fit the signal plus background contribution to the data, and extract a 95% C.L. upper limit on the possible signal contribution. The obtained results are plotted as a dotted line in Fig. 1. This temporal analysis relies only on the flatness of the background shape; it does not have the counting experiment's systematic uncertainty on the background normalization. Consequently, its dominant systematic uncertainty is the 11% uncertainty on the luminosity measurement. For a mass difference $m_{\tilde{g}} - m_{\tilde{\chi}_1^0} > 100$ GeV$/c^2$, assuming BR$(\tilde{g} \to g\tilde{\chi}_1^0) = 100\%$, we are able to exclude $m_{\tilde{g}} < 382$ GeV$/c^2$ at the 95% C.L. for a lifetime of 10 $\mu$s with the time-profile analysis.

We have presented the results of the first search for long-lived gluinos produced in 7 TeV $pp$ collisions at the LHC. We looked for the subsequent decay of those gluinos that would have stopped in the CMS detector during time intervals where there were no $pp$ collisions. In particular, we searched for decays during gaps in the LHC beam structure. We recorded such decays with dedicated calorimeter triggers. In a dataset with a peak instantaneous luminosity of $1 \times 10^{32}$ cm$^{-2}$s$^{-1}$, an integrated luminosity of 10 pb$^{-1}$, and a search interval corresponding to 62 hours of LHC operation, no significant excess above background was observed. Limits at the 95% C.L. on gluino pair production over 13 orders of magnitude of gluino lifetime are set. For a mass difference $m_{\tilde{g}} - m_{\tilde{\chi}_1^0} > 100$ GeV$/c^2$, assuming BR$(\tilde{g} \to g\tilde{\chi}_1^0) = 100\%$, we exclude $m_{\tilde{g}} < 370$ GeV$/c^2$ for lifetimes from 10 $\mu$s to 1000 s with a counting experiment. Under the same assumptions, we are able to further exclude $m_{\tilde{g}} < 382$ GeV$/c^2$ at the 95% C.L. for a lifetime of 10 $\mu$s with a time-profile analysis. These limits are the most restrictive to date.

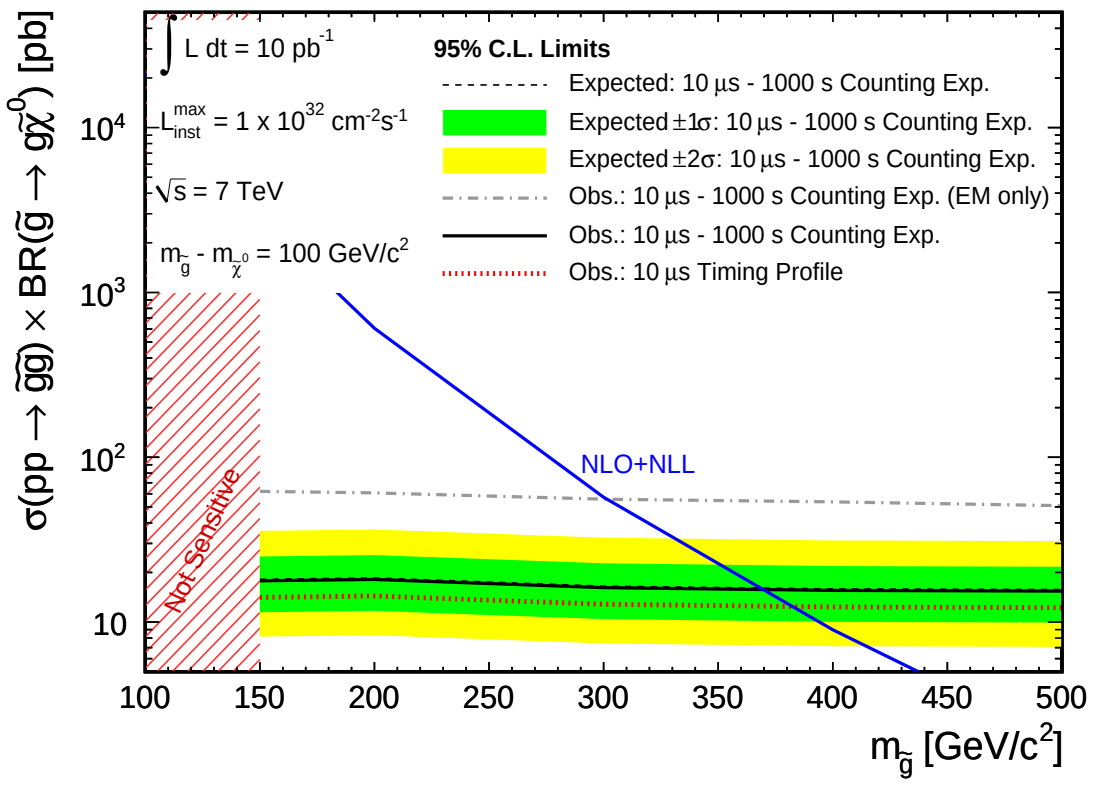

***Fig. 2:** 95% C.L. limits on gluino pair production cross section times branching fraction as a function of gluino mass assuming the "cloud model" of R-hadron interactions (solid line) and EM interactions only (dot-dashed line). The $m_{\tilde{g}} - m_{\tilde{\chi}_1^0}$ mass difference is maintained at 100 GeV$/c^2$; results are only presented for $m_{\tilde{\chi}_1^0} > 50$ GeV$/c^2$.*

The author gratefully acknowledges the support by the Helmholtz Alliance "Physics at the Terascale".

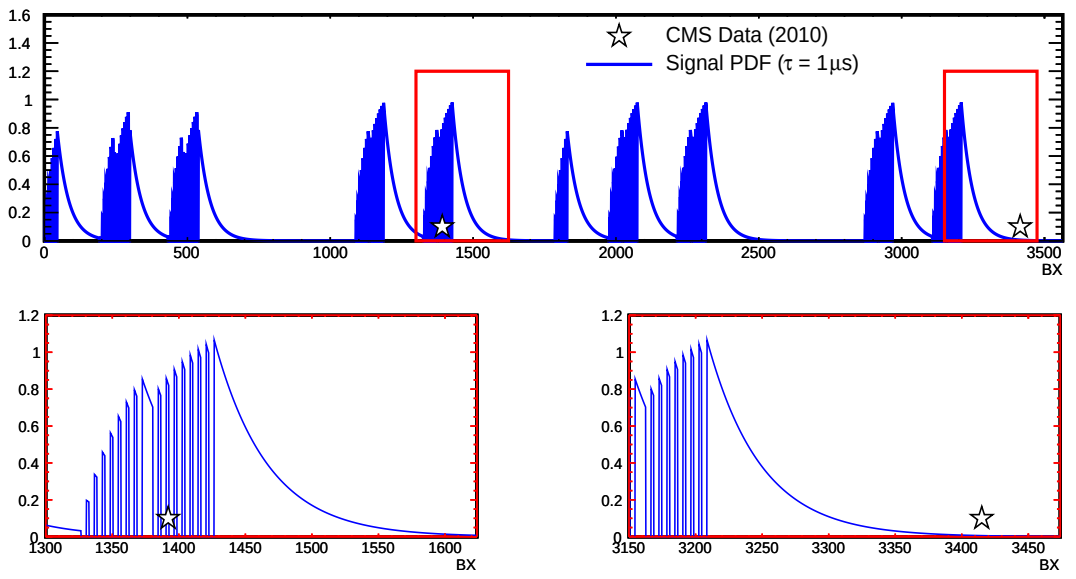

***Fig. 3:*** *The top panel shows the in-orbit positions of 2 observed events in the subset of our data that was recorded during an LHC fill with 140 colliding bunches. The decay profile for a 1 $\mu$s lifetime hypothesis is overlaid. The bottom panels are zoomed views of the boxed regions around the 2 events in the top panel so that the exponential decay shape of the signal hypothesis can be seen.*